\documentclass[twocolumn,showpacs,preprintnumbers,amsmath,amssymb,floatfix]{revtex4}
\usepackage{ifpdf}
\usepackage[dvips]{color} 
\usepackage[latin1]{inputenc}
\usepackage{SIunits}

\ifpdf
	\usepackage[pdftex]{graphicx}
\else
	\usepackage{graphicx}
	\usepackage[colorlinks=false]{hyperref}
\fi

\begin{document}
\draft
\title{N,N'-dimethylperylene-3,4,9,10-bis(dicarboximide) on alkali halide(001) surfaces}
\author{Markus Fendrich\footnote{email: markus.fendrich@uni-due.de}, Manfred Lange, Christian Weiss, Tobias Kunstmann and Rolf Möller} 
\affiliation{
Fachbereich Physik, Universität Duisburg-Essen, D-47048 Duisburg, Germany}

\begin{abstract}
The growth of N,N'-dimethylperylene-3,4,9,10-bis(dicarboximide) (DiMe-PTCDI) on KBr(001) and NaCl(001) surfaces has been studied. Experimental results have been achieved using frequency modulation atomic force microscopy at room temperature under ultra-high vacuum conditions. On both substrates, DiMe-PTCDI forms molecular wires with a width of 10 nm, typically, and a length of up to 600 nm at low coverages. All wires grow along the [110] direction (or [1$\bar{1}$0] direction, respectively) of the alkali halide (001) substrates. There is no wetting layer of molecules: Atomic resolution of the substrates can be achieved between the wires. The wires are mobile on KBr surface but substantially more stable on NaCl. A p(2 x 2) superstructure in brickwall arrangement on the ionic crystal surfaces is proposed based on electrostatic considerations. Calculations and Monte-Carlo simulations using empirical potentials reveal possible growth mechanisms for molecules within the first layer for both substrates, also showing a significantly higher binding energy for NaCl(001). For KBr, the p(2 x 2) superstructure is confirmed by the simulations, for NaCl, a less dense, incommensurate superstructure is predicted. 
\end{abstract}

\maketitle

\section{Introduction}
The possible use of organic molecules for advanced types of electronic devices has attracted much interest in recent decades.\cite{Dimitrakopoulos2002,Horowitz1998,jaeckel04PRL92_188303} A large field of research has emerged, studying the structural and electronic properties of various molecules on solid surfaces. Scanning probe methods have proved to be a powerful tool for these studies, especially scanning tunneling microscopy (STM) has been used for many promising studies.\cite{moresco01PRL86_672,Temirov06Nature444_350,Liljeroth07Science317_1203} Nevertheless, STM is limited to conducting surfaces, while insulating surfaces are still important: Here, the influence of the substrate on the electronic properties is small compared to metal or semiconductor surfaces.
\par Frequency modulation atomic force microscopy (FM-AFM)\cite{Albrecht91JAP69_668,giessibl95Science267_68} is a scanning probe technique that allows for real space imaging of insulating surfaces. Thus it has made the analysis of the structural properties of organic molecules on insulators accessible.\cite{nony03nano15_S91,nony04NanoL4_2185} Recent studies demonstrate the capability of the technique to analyse the growth of the prototypical semiconductor molecules C$_{60}$\cite{Burke2007,Burke05PRL94_096102} and 3,4,9,10-perylene-tetracarboxylic-dianhydride (PTCDA)\cite{Fendrich07Nano18_084004,Kunstmann05PRB71_121403(R),Mativetsky07Nanotechnology18_105303,Dienel2008,Burke2008} on alkali halide (001) surfaces.
\par However, all these studies show dewetting of the surfaces and the growth of bulk-like molecular structures for PTCDA or fractal islands for C$_{60}$. For the development of small-scale electronic devices, a controlled growth of low-dimensional molecular structures would be desirable.\cite{Schnadt2008} This kind of growth has recently been analysed for a perylene derivate similar to PTCDA, N,N'-dimethylperylene-3,4,9,10-bis(dicarboximide) (DiMe-PTCDI); \cite{Gavrila04APL85_4657,Hadicke86ACC42_189,Schaefer01ADFM11_193,Schaefer98TSF379_176} this molecule shows the growth of wire-like structures on KBr(001).\cite{Fendrich07APL91_023101} 
\par The present paper reviews the results for KBr(001) and shows experimental data for another alkali halide surface, NaCl(001), as well as molecular force field calculations analysing the molecular arrangement of DiMe-PTCDI on both  substrates.

\section{Experimental}
The experiments have been performed at room temperature under ultra-high vacuum(UHV) conditions in a commercially available analysis system (AFM/STM by Omicron, Germany) operated in FM-AFM mode. The oscillation of the cantilever is detected by beam deflection; to illuminate the cantilever, a SuperLum LED coupled to the UHV chamber by a single-mode fiber was used. For the detection of the frequency and control of the cantilever oscillation amplitude, we used the easyPLLplus system (Nanosurf, Switzerland), operating in self oscillation mode. The specifications of the silicon cantilevers (PPP-NCL by Nanosensors, Switzerland) were: resonant frequencies of $f_0\approx 160$ kHz, spring constants of $k\approx$ 30 - 40 N/m and quality factors of $Q\approx$ 30000 in UHV.  The cantilevers were driven at an amplitude of $A = 20$ nm, typically. In some experiments, a voltage between tip and sample was applied to compensate for electrostatic interactions. Specific frequency shifts are given for each image. 
\par For scanning control and data acquistion, the open-source software GXSM\cite{Zahl03RSI74_1222} combined with home-built electronics has been used. Image processing has been done using the free software WSXM.\cite{wsxm07}
\par Clean KBr and NaCl (001) surfaces have been prepared by \textit{in situ} cleaving of commercially available single crystals (Korth, Germany). The crystals have been heated at 450 K for 1 h before cleaving to remove contaminations and also after cleaving to remove trapped charges. After preparation of the substrates, various amounts of DiMe-PTCDI have been evaporated onto the samples from a home-built crucible. The temperature of the oven was 550 K, the evaporation rate was checked before and after deposition by a quadrupole mass spectrometer. During evaporation, the sample was kept at room temperature.

\section{Experimental results}

\subsection{KBr(001)}
\begin{figure}[h!]
\includegraphics[width=8.6cm]{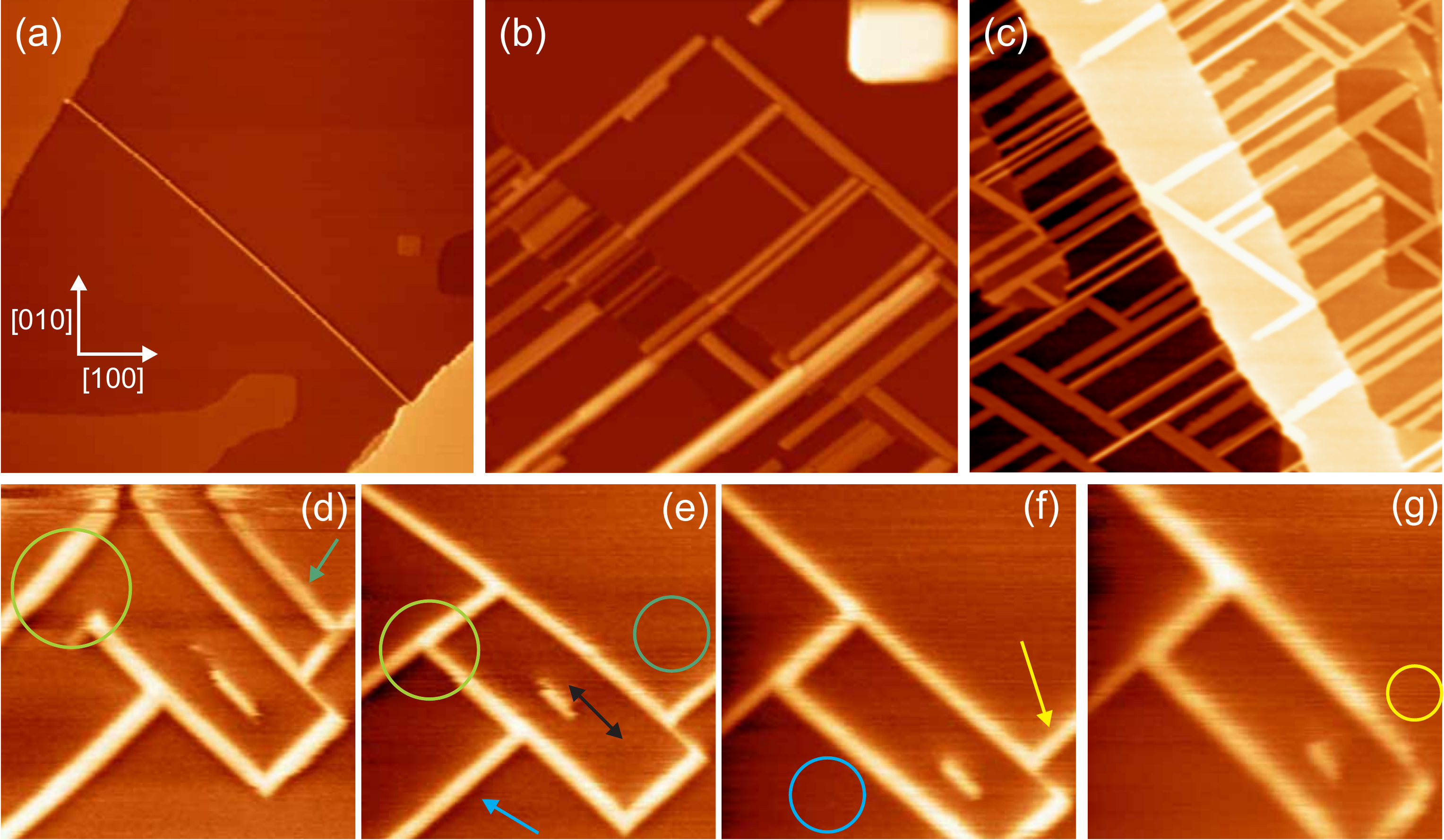}
\caption{(color online) \cite{wsxm07} Review of the previously reported experimental results for DiMe-PTCDI on KBr(001).\cite{Fendrich07APL91_023101} The organic molecules grows wire-like structures with height of at least 2 molecular layers at low (a) and higher (b) coverages, all wires aligned with $\left\langle 110 \right\rangle$ surface axes. When the sample is cooled during evaporation (c), the height of the wires is only one layer. The wires are stabilized by substrate step edges; when no step edges are present, the wires are unstable and diffuse, as shown in the series of images (d)-(g). All images have been obtained at small frequency shifts, df = -0.5 Hz., image sizes are (a) 600 $\times$ 600 nm$^2$, (b) and (c) 400 $\times$ 400 nm$^2$, (d)-(g) 300 $\times$ 300 nm$^2$. }
\label{kbr_old}
\end{figure}

On KBr(001) surfaces, DiMe-PTCDI grows wire-like islands with a length of up to 600 nm (see Fig. \ref{kbr_old}). Between the wires, atomic resolution of the KBr substrate could be achieved: There is no wetting layer of organic molecules. All wires are aligned with $\left\langle 110 \right\rangle$ surface axes. The wires are stabilized by step edges, ''loose ends'' are seldom observed. When no step edges are present, the structures are highly unstable and diffuse when scanned for a longer time. The wires have a height of at least two molecular layers.
\par Due to their instability, no molecular resolution of the wires was possible, so no information on the arrangement of the molecules within the wire, could be achieved experimentally. For the initial growth, i.e., for the first layer upon the KBr substrate, a model proposing a p(2$\times$2) superstructure in brickwall arrangement has been elaborated. This model was based on force field calculations and electrostatic considerations for a single molecule on the ionic KBr(001) surface.\cite{Fendrich07APL91_023101}

\subsection{NaCl(001)}

\begin{figure}[h!]
\includegraphics[width=8.6cm]{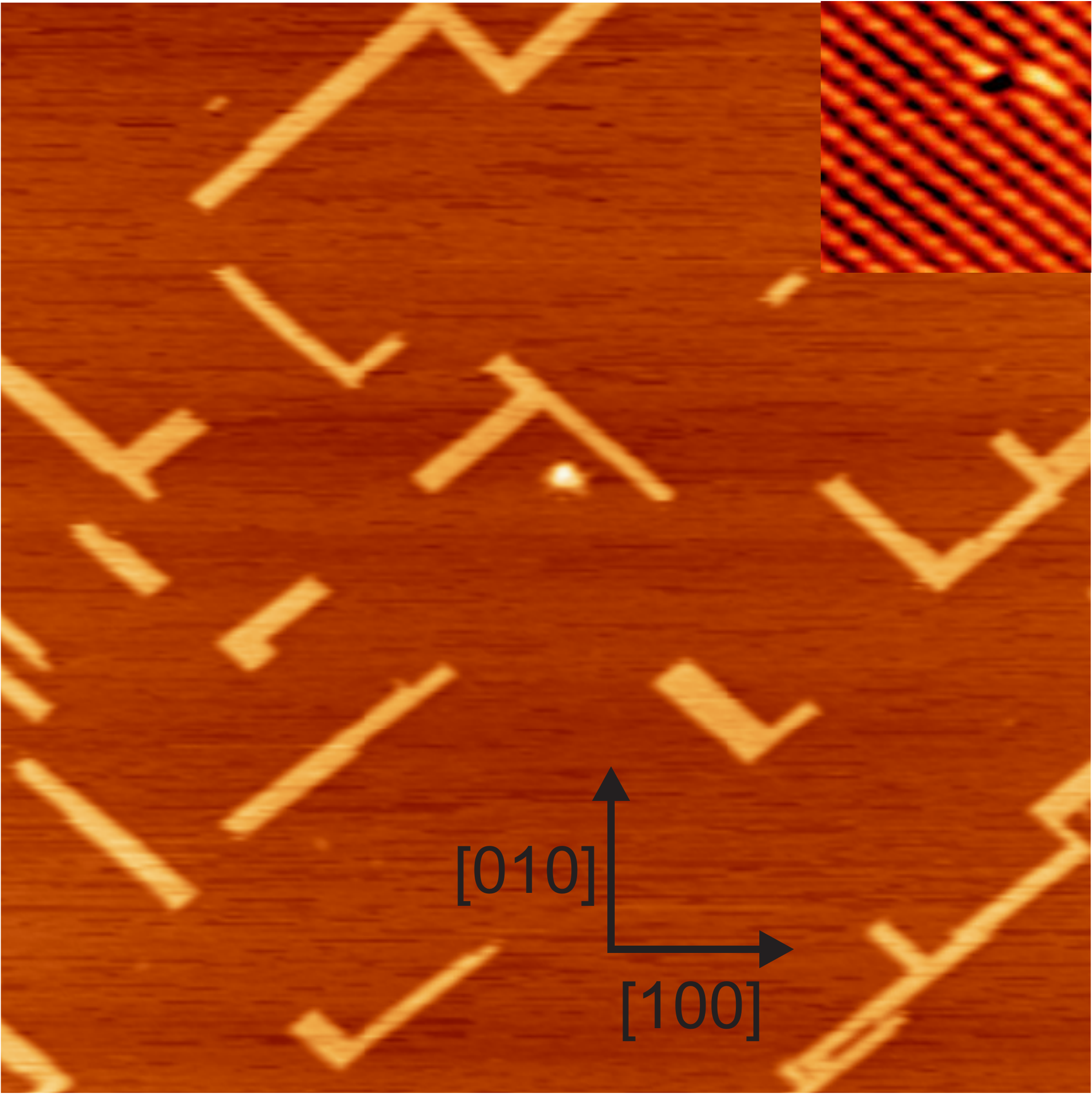}
\caption{(color online) \cite{wsxm07} Molecular wires of DiME-PTCDI on NaCl(001) at low coverage (0.2 ML). The image has been obtained at a frequency shift of df = -0.5 Hz, the size of the image is 400 $\times$ 400 nm$^2$. No NaCl step edges can be seen within the image. All wires are oriented towards $\left\langle 110 \right\rangle$ surface directions and very stable. No changes could be observed while scanning this area for several hours. There is no wetting layer of molecules, between the wires, atomic resolution of the substrate is possible (see inset, 3 $\times$ 3 nm$^2$, df = -6 Hz).}
\label{nacl_low}
\end{figure}

On NaCl(001) surfaces, DiMe-PTCDI molecules show similar structures. As shown in Fig. \ref{nacl_low}, wires are still aligned with $\left\langle 110 \right\rangle$ surface axes. No wetting layer of molecules is found. In contrast to KBr, the wires are much more stable, even if they are not stabilized by step edges: The area shown in the image has been scanned for several hours without showing any changes in the arrangement of the molecular islands. All the wires are two molecular layers high. 
\par On KBr, continuous networks of wires have been found, supposably due to stability reasons; here, we find smaller islands that are not interconnected and sometimes consist of several merged perpendicular wires. No step edges are needed as nucleation centers for the islands on NaCl; it seems reasonable that islands nucleate at surface defects.

\par
\begin{figure}[h!]
\includegraphics[width=8.6cm]{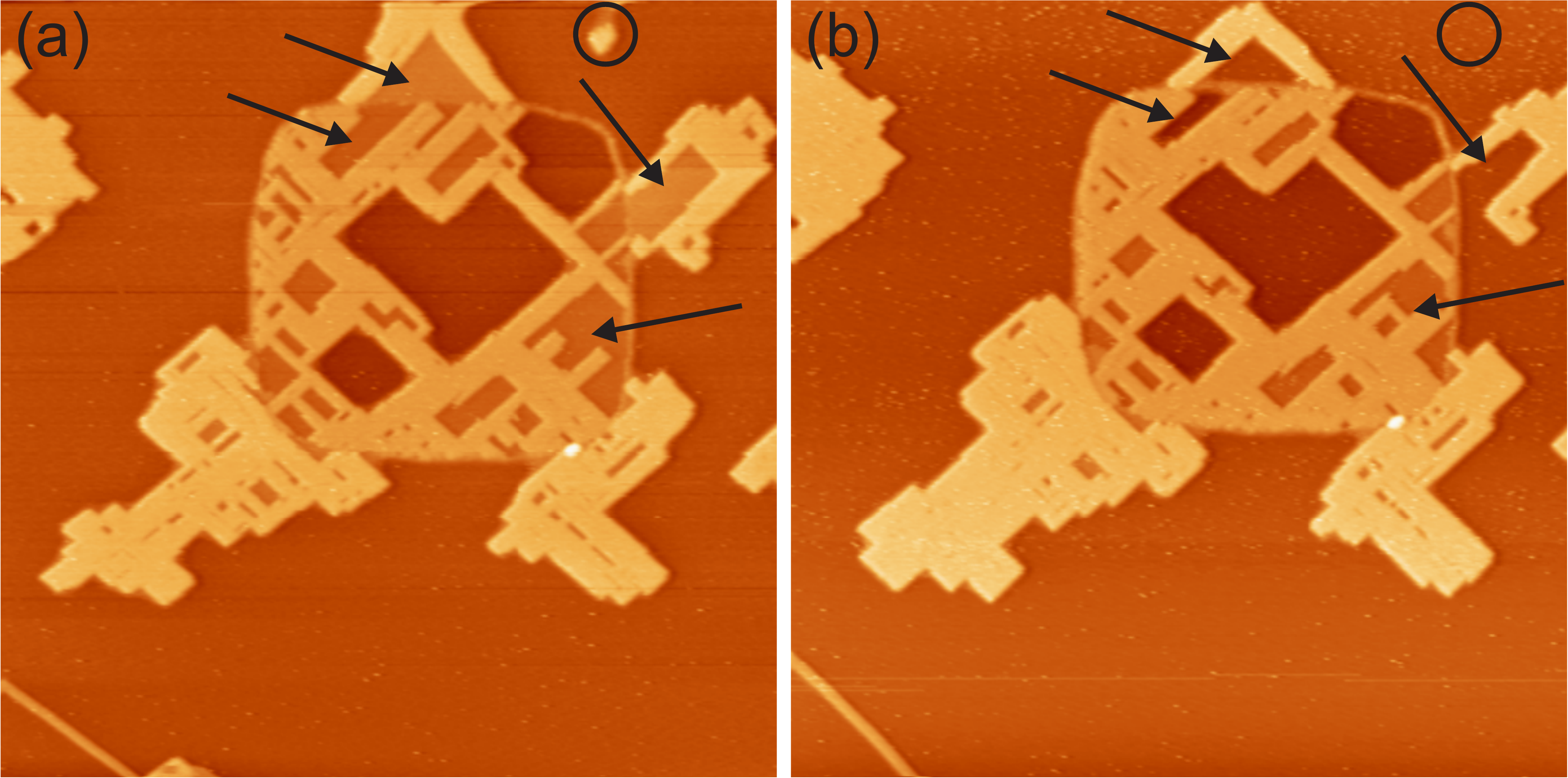}
\caption{(color online) \cite{wsxm07} Images (a) and (b) show two subsequent scans of a 500 $\times$ 500 nm$^2$ large area at high coverage (the total coverage is about 2 ML of DiMe-PTCDI). The time between (a) and (b) is about 30 min. One can see large islands of molecules with heights between 1 and 4 molecular layers. The rectangular structure aligned with $\left\langle 110 \right\rangle$ surface axes can still be seen, in contrast to the situation at low coverages, the islands are more compact and do not show wire-like growth. The ML-covered areas (see arrows) are not stable, the dewetting of these areas and some changes in the 
shapes of the higher areas can be seen in (b). Both images have been obtained at a frequency shift of df = -0.5 Hz.}
\label{dewetting}
\end{figure}

 When the coverage of molecules on the surface is increased, larger and more compact molecular islands are found while the framework of the perpendicularly arranged wires can still be seen. Islands with a height of up to 4 molecular layers have been observed. Within these islands, 1-layer areas are embedded; however, these wetting layers are unstable, a dewetting of the 1-layer areas is observed within several minutes. Fig. \ref{dewetting} shows two subsequent scans of the same area (500 $\times$ 500 nm$^2$). One can see the molecules dissappear from the enclosed areas covered with only one layer of molecules. These molecules diffuse two higher layers, as reported also for C$_{60}$ molecules on KBr(001).\cite{Burke2007,Burke05PRL94_096102} 
\par
\begin{figure}[h!]
\includegraphics[width=8.6cm]{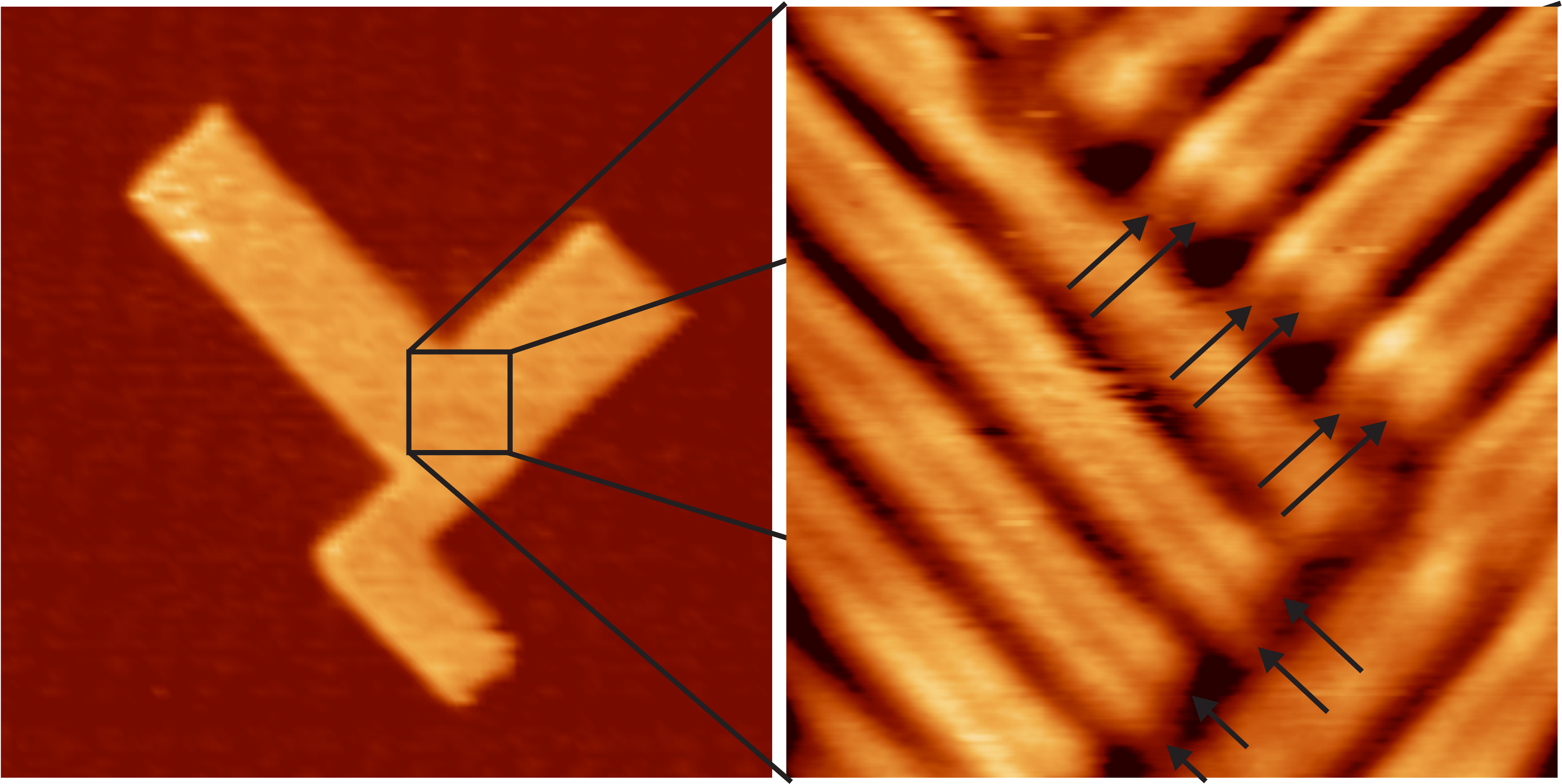}
\caption{(color online) \cite{wsxm07} The left image (100 $\times$ 100 nm$^2$, df = -0.5 Hz) shows a 2-layer high island of DiMe-PTCDI on NaCl(001). The island is merged out of several perpendicular wires, the right image shows a zoom (12 $\times$ 12 nm$^2$, df = -5.5 Hz) of the junction. Molecular rows can be resolved; at the junction, every second row appears shorter (see arrows). Therefore, a brickwall arrangement of the molecules seems reasonable.}
\label{2ndlayer}
\end{figure}

 As the islands are wider and more stable than on KBr(001), molecular resolution of the topmost layer of DiMe-PTCDI on NaCl(001) was possible. Fig.\ref{2ndlayer} shows resolution of molecular rows on top of an island which is 2 layers high and merged out of 3 wires. The junction can clearly be seen on the molecular scale as an interface of perpendicular rows. The internal structure of the molecular rows cannot be resolved; nevertheless, every second row seems shorter at the junction, thus a brickwall arrangement of the molecules seems reasonable.

\section{Calculations}

To get further insight into the growth mechanisms for DiMe-PTCDI on alkali halide (001) surfaces and corrobarate the models based on the experimental findings, molecular force field calculations based on empirical potentials have been performed. The capability of empirical potentials has been shown for several molecular systems.\cite{Mannsfeld04PRB69_075461,Mannsfeld05PRL94_056104,Fendrich06PRB73_115433,Fendrich07PRB76_121302(R)} In particular, similar calculations could elucidate the experimental data for a very similar system, PTCDA on KCl(001).\cite{Dienel2008}
\par As preliminary calculations, the potential energies of single DiMe-PTCDI molecule on KBr(001) and NaCl(001) have been calculated using the empirical potentials of the AMBER molecular force field.\cite{Weiner84JACS106_765,Weiner86JCC7_230,Cornell95JACS117_5179} As DiMe-PTCDI is a polar molecule (the electronegativity of the 4 oxygen atoms causes a shift of negative charge to the ''corners'' of the molecule), the electrostatics of the system have to be accounted for. The polarity of the molecule is represented by partial charges assigned to each atom within the molecule by a Mulliken method, using the software Hyperchem. Each ion of the substrate is represented by a point charge of $\pm e_0$. 
\par
\begin{figure}[h!]
\includegraphics[width=8.6cm]{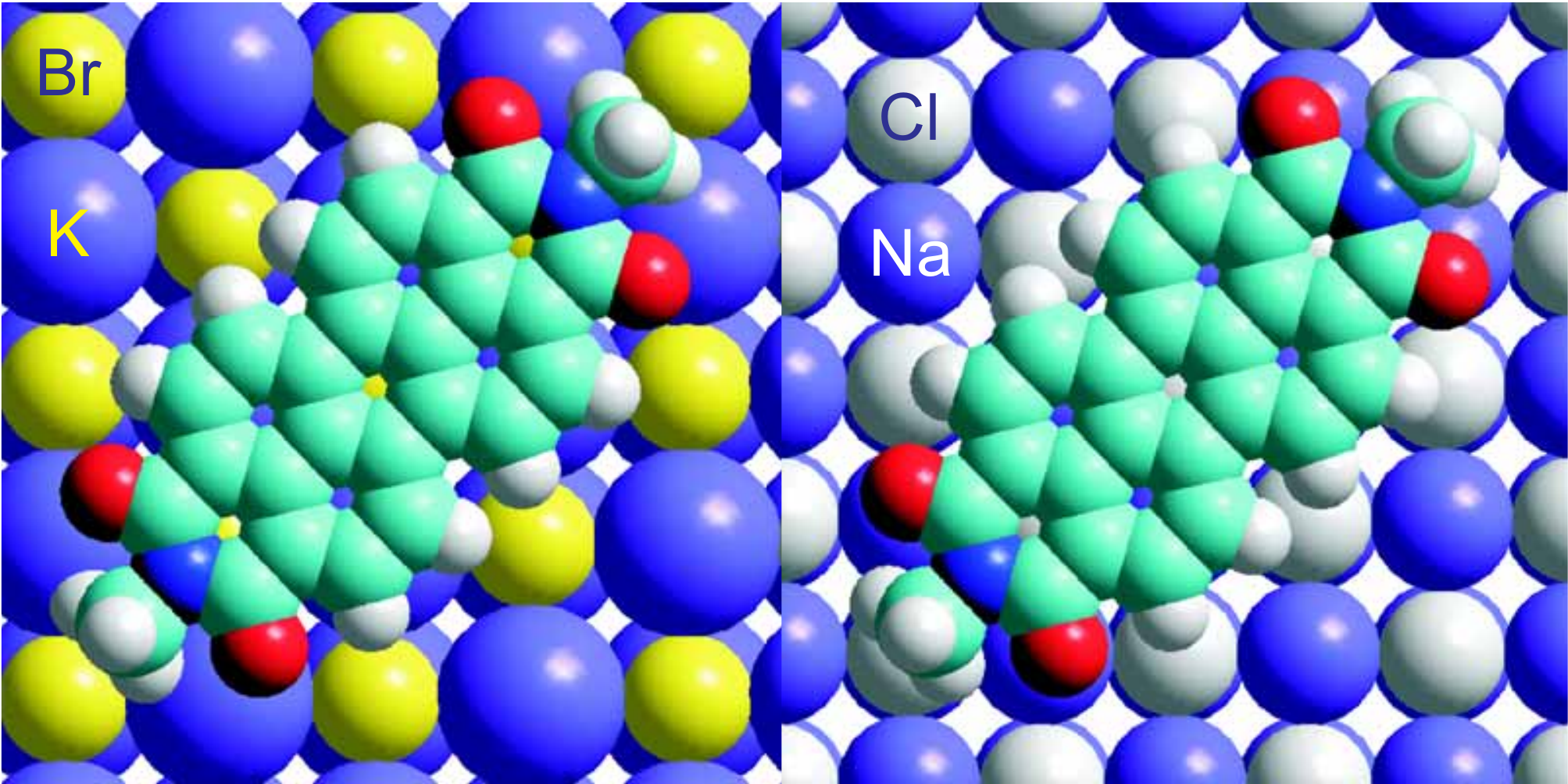}
\caption{(color online) Minimum energy positions for DiMe-PTCDI molecules on KBr(001) and NaCl(001), calculated with an AMBER force field calculation. In both cases, the minimum energy position is aligned with the [110] substrate axis, on top of a halide ion (Br$^-$ or Cl$^-$). Due to the smaller lattice constant of NaCl, the binding energy is higher for this substrate($E_{NaCl}$ = -1.65 eV, $E_{KBr}$ = -1.35 eV).}
\label{modell_beide}
\end{figure}

 Calculating the potential energy as described above, the minimum energy position for a single molecule was determined for both substrates using a gradient search algorithm. Optimum positions are found as shown in Fig. \ref{modell_beide}, i.e., on top of a halide ion, the molecule aligned with the [110] axis of the substrate. In this way, the negatively charged oxygen atoms are close to 4 positive alkali ions of the substrate. The adsorption geometry is more favorable for NaCl, which has a smaller lattice constant of 562 pm compared to KBr (658 pm); therefore, the binding energy of a single molecule is higher for NaCl ($E_{NaCl}$ = - 1.65 eV, $E_{KBr}$ = -1.35 eV). 

\par In a second step, potential energies for larger molecular domains have been calculated. The Grid method and the software Powergrid \cite{Mannsfeld04Diss,Mannsfeld04PRB69_075461} have been used to calculate the potential energy for domains consisting of 150 molecules or more. The molecular arrangement within the domain has been optimized using a Monte-Carlo method, the Metropolis algorithm.\cite{Metropolis53JCP21_1087} Several Monte-Carlo experiments have been performed: After the Metropolis algorithm was started for an arbitrary initial configuration, the arrangement was optimized, i.e. the size of the unit cell and the orientation of the molecules within the unit cell were changed until an optimum arrangement was found. The calculation confirms the supposed p(2 $\times$ 2) brickwall arrangement for one layer of DiMe-PTCDI molecules on KBr(001). For NaCl(001), a incommensurate structure with an epitaxy matrix of 
\begin{eqnarray}\notag
	C = \left( \begin{array}{cc} 
	2.08 & 0.08\\
	0.08 & 2.08
	\end{array} \right)
\end{eqnarray}
was found. Within the domain, all the molecules are still aligned towards the $\left\langle 110 \right\rangle$ substrate axis.
\par The higher stability of the islands on NaCl(001) reflects in a higher binding energy for the domains. The energy per molecule within a domain is $E_{KBr,dom}$ = -1.4 eV for KBr and $E_{NaCl,dom}$ = -1.8 eV for NaCl. Table \ref{kbr_vs_nacl} shows the results of the calculations.

\begin{table}[h!]
\caption{Summary of the results of the Monte Carlo simulations for DiMe-PTCDI on KBr(001) and NaCl(001).}
\vspace{0.5cm}
\begin{tabular}{l||c|c}
  & KBr(001) & NaCl(001)\\
  \hline
  structure & commensurate & incommensurate\\
  &p(2 $\times$ 2) & \\
  & brickwall& brickwall\\
  \hline
  area per& A = 1.74 nm$^2$ & A = 1.37 nm$^2$\\
  molecule&&\\
  \hline
  binding energy & -1.35 eV (single)&-1.65 eV(single)\\
  per molecule &-1.4 eV (domain)& -1.8 eV (domain)\\
\end{tabular}
\label{kbr_vs_nacl}
\end{table}

\section{Conclusion}
We have shown an experimental study of DiMe-PTCDI on two alkali halide (001) substrates, KBr(001) and NaCl(001). On both substrates, we have observed the growth of wire-like molecular islands which are oriented along the $\left\langle 110 \right\rangle$ substrate axis. The wires are at least 2 molecular layers high for both substrates. The wires are very mobile on KBr surfaces, where they are stabilized by step edges; on NaCl surfaces, the wires are more stable, allowing for the resolution of molecular rows within the topmost molecular layer. When larger amounts of molecules are evaporated onto NaCl(001) surfaces, large, compact islands are found. In contrast to the two-layer wires at low coverages, monolayer-covered areas are found. However, a dewetting of these areas was observed, possibly induced or supported by the scanning tip.
\par Calculations based on empirical potentials give further insight to the structural properties. For KBr(001), a p(2 $\times$ 2) superstructure is found, all molecules are aligned towards the $\left\langle 110 \right\rangle$ substrate axis. This alignment is also found for NaCl(001), however, in this case an incommensurate superstructure is predicted. However, at least this purely theoretical finding has to be doubted, as there are no experimental hints. Inaccuracies of the calculation may occur for several reason: In contrast to the experiment, only one layer of molecules on the substrates is considered; all molecules are lying flat on the surface, a constriction that is necessary for the Grid method;\cite{Mannsfeld04PRB69_075461} and possible conformational changes of molecules or substrates are completely neglected, both molecules and substrates are rigid. In spite of the inaccuracies of the calculations, the higher stability for NaCl reflects in a higher binding energy for this substrate compared to KBr.

\section{Acknowledgement}
Financial support is granted by the Deutsche Foschungsgemeinschaft through SFB616 ''Energy Dissipation at Surfaces''.

\end{document}